\documentclass[conference]{IEEEtran}
\IEEEoverridecommandlockouts

\usepackage{cite}
\usepackage{amsmath,amssymb,amsfonts}
\usepackage{graphicx}
\usepackage{textcomp}
\usepackage{xcolor}
\usepackage{booktabs}
\usepackage{url}
\usepackage{tikz}
\usetikzlibrary{positioning,arrows.meta,fit,backgrounds}

\begin{document}
\bstctlcite{paper:reference-style}

\title{Cyber-JEV: Matched Contrastive Fine-Tuning of Language Models for Fast Source-Code Vulnerability Triage}

\author{\IEEEauthorblockN{Zhuyi Zhao\textsuperscript{*}, Shengjie Yuan\textsuperscript{*}, Shengyu Wang, Kang Wei,\\
Wang Yang, Wenchao Xu, Fushuo Huo}
\thanks{\textsuperscript{*}Zhuyi Zhao and Shengjie Yuan contributed equally to this work and are co-first authors.}}

\maketitle

\begin{abstract}
Single-pass scoring models such as Jev give each code function a vulnerability probability, so that security teams can
rank many functions for review without generating written explanations. Training an open-weight model for this task is
limited by data. Benchmarks with both vulnerable and non-vulnerable labels keep few vulnerable functions once examples
related to the test data are removed, while larger public vulnerability records mostly contain vulnerable functions
only. Mixing the two lets the data source predict the label, and training on the labeled data alone leaves a model that
often scores a fixed function above its vulnerable original. We present Cyber-JEV, a fine-tuning method that uses fix
commits to turn these records into comparison groups. Each group holds a vulnerable function, its patched version and
up to two unchanged functions from the same file, so the model sees the change that removed a recorded defect. One
shared model learns two tasks: classifying functions from the labeled reference data, and ranking the vulnerable
function above the other members of its group. At test time the input is one function and the output is a
vulnerability score. On a test benchmark of 7,270 functions, an ensemble of five fine-tuned Qwen3.5-4B models reaches
an area under the receiver operating characteristic curve of 0.814, compared with 0.743 for Jev 1.13. With the 4,031
alerts of Jev as the review budget, it misses 16 of 322 vulnerable functions instead of 39. Compared with training on
the reference data alone, it ranks the vulnerable version above the patched version 14.8 to 29.2 percentage points more
often on three sets of real fixes. Such scores can order the changed functions of a pull request or a release for
review under a fixed budget.
\end{abstract}


\section{Introduction}

Modern continuous integration and pre-release security auditing workflows must screen thousands of modified functions under tight latency and analyst review budgets. Rather than relying on slow multi-turn chain-of-thought generation, fast probabilistic decision language models---exemplified by TypeSafe AI's Jev~\cite{typesafe2026jev,typesafe2026api}---provide a streamlined single-pass interface: given a candidate function, the model outputs a calibrated vulnerability probability in a single forward pass, allowing automated pipelines to prioritize high-risk functions or trigger human review.

Extending this fast decision paradigm to source-code vulnerability triage, however, faces two competing training bottlenecks~\cite{steenhoek2023empirical,ullah2024llmsec,risse2024limits}. The first bottleneck is \emph{source shortcut learning} when aggregating public security data. After standard cross-project and near-duplicate deduplication hygiene is enforced between training and evaluation splits~\cite{allamanis2019duplication,croft2023dataquality,ding2025primevul,kapoor2023leakage}, clean benchmark pools retain very few vulnerable positives (only 197 functions in our setup). Augmenting training with public security patch archives such as MoreFixes~\cite{akhoundali2024morefixes} and MegaVul~\cite{ni2024megavul} contributes thousands of real vulnerabilities, yet the records we admit from these archives contain only vulnerable functions and no benign negatives. Naive joint training allows repository provenance to give away the label: a classifier can minimize training loss simply by recognizing project formatting and identifier conventions, causing test area under the receiver operating characteristic curve (AUROC) to drop to 0.602---worse than ranking functions purely by byte length.

\begin{figure*}[t]
\centering
\begin{tikzpicture}[font=\footnotesize,
  box/.style={draw, rounded corners=2pt, align=center, inner sep=2.5pt},
  arr/.style={-{Latex[length=1.6mm]}, semithick}]
\node[font=\footnotesize\itshape] at (2.45,2.45) {Data roles};
\node[font=\footnotesize\itshape] at (7.9,2.45) {Groups built from fix commits};
\node[font=\footnotesize\itshape] at (12.9,2.45) {Training terms};
\node[font=\footnotesize\itshape] at (16.55,2.45) {Use};
\node[box, fill=blue!7, text width=4.7cm] (ref) at (2.45,1.6) {Reference training data\\PrimeVul, DiverseVul, CVEfixes\\197 vulnerable, 5,026 not vulnerable};
\node[box, fill=orange!12, text width=4.7cm] (ext) at (2.45,0.25) {External vulnerability records\\17 source labels, e.g., MoreFixes\\3,357 vulnerable, no other labels};
\node[box, dashed, fill=gray!6, text width=4.7cm] (eval) at (2.45,-1.1) {Evaluation data, never trained on\\test 7,270; dev.\ and calib.\ 512;\\TitanVul pair sets A72, P309};
\node[box, fill=gray!10, text width=4.2cm] (fix) at (7.9,0.75) {Fix commit\\verified files before and after,\\complete functions extracted};
\node[box, fill=red!12, text width=3.9cm] (vul) at (7.9,-0.3) {vulnerable function};
\node[box, fill=blue!7, text width=3.9cm, below=1.2mm of vul] (pat) {patched version};
\node[box, fill=blue!7, text width=3.9cm, below=1.2mm of pat] (unc) {$\le$\,2 unchanged functions, same file};
\node[draw, dashed, rounded corners=2pt, fit=(vul)(pat)(unc), inner sep=2.5pt,
  label={[font=\scriptsize]below:2,829 groups; 725 vulnerable functions have no group}] (grp) {};
\node[box, fill=green!9, text width=3.0cm] (cls) at (12.9,1.6) {Classification term\\reference data only,\\both labels};
\node[box, fill=green!9, text width=3.0cm] (rank) at (12.9,-0.6) {Ranking term\\vulnerable member above\\the rest of its group};
\node[box, fill=yellow!18, text width=2.5cm] (model) at (16.55,0.5) {One shared model\\input: one function\\output: a score};
\draw[arr] (ref.east) -- node[above, font=\scriptsize] {labels} (cls.west);
\draw[arr] (ref.south east) -- ([yshift=2pt]fix.west);
\draw[arr] (ext.east) -- ([yshift=-2pt]fix.west);
\draw[arr] (fix.south) -- (vul.north);
\draw[arr] (grp.east) -- (rank.west);
\draw[arr] (cls.east) -- (model.north west);
\draw[arr] (rank.east) -- (model.south west);
\end{tikzpicture}
\caption{Cyber-JEV workflow. Reference data train the classification term; groups built from fix commits train only
the ranking term. Both terms update one model, which scores a single function at test time.}
\label{fig:overview}
\end{figure*}
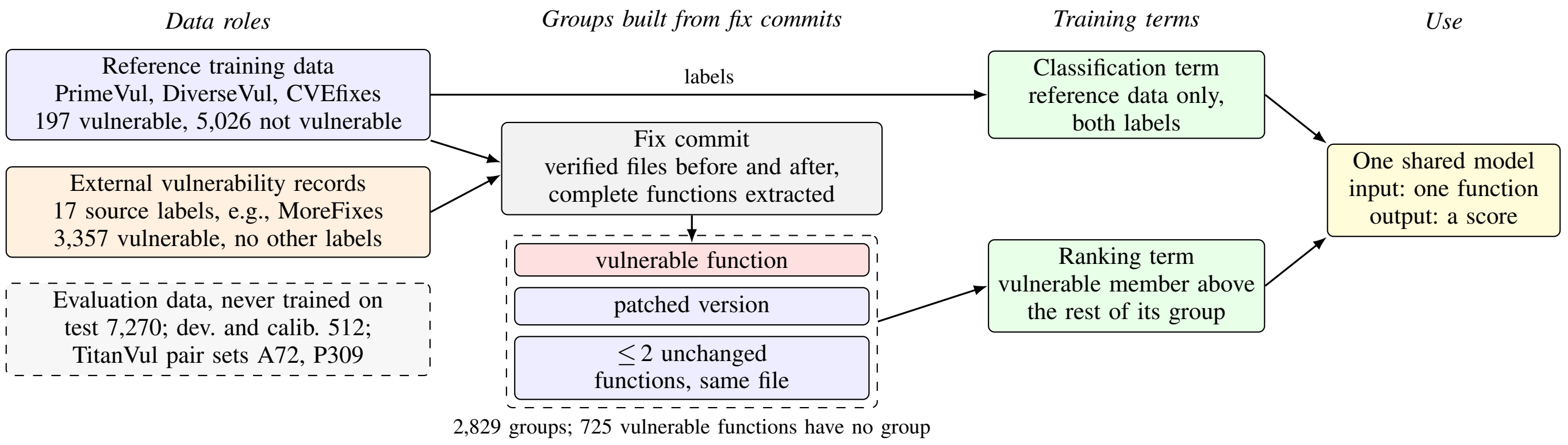

The second bottleneck is an inherent \emph{anti-patch bias}. If external patch archives are simply discarded in favor of standard binary cross-entropy fine-tuning on the clean reference pool, the model attains strong macroscopic ranking (0.804 AUROC) but systematically fails on minimal security fixes: in 67.2\% of vulnerability-versus-patch pairs, it assigns a \emph{higher} vulnerability score to the patched code than to the original vulnerable code (mirroring the 60.5\% inversion rate of the unadapted base model). A likely reason is that security patches typically insert bounds checks, null-pointer guards, or error-return branches, so standard cross-entropy may teach the model to flag high-risk code modules broadly while mistaking newly added defensive checks for risk indicators.

Existing distributionally robust optimization~\cite{sagawa2020groupdro}, invariant risk minimization~\cite{arjovsky2019irm}, and prior-shift logit adjustment~\cite{menon2021logit} all require both positive and negative examples within every source, and thus can neither construct missing within-source negatives nor teach fine-grained patch discrimination. Our approach draws inspiration from classical matched statistical designs and modern contrastive learning: matched sampling in statistics (such as classical case-control studies~\cite{breslow1980casecontrol}) eliminates shared group bias via conditional likelihood~\cite{neyman1948incidental,andersen1970cml} while relying on a representative reference sample to anchor baseline probabilities~\cite{prentice1979casecontrol,scott1986choice}; meanwhile, counterfactual minimal edits~\cite{kaushik2020cad} and hard-negative contrastive ranking~\cite{oord2018cpc,robinson2021hard,liu2023contrabert} directly train neural encoders to detect subtle semantic changes.

Guided by these insights, we introduce Cyber-JEV, adapting the open-weight Qwen3.5-4B model~\cite{qwen35report} for fast, calibrated, and patch-sensitive source-code vulnerability triage (Fig.~\ref{fig:overview}). At the data level, we develop Provenance-Matched Completion (PMC), a five-stage curation and sanitization pipeline that extracts dual-hash and abstract syntax tree (AST) verified patched versions and unchanged same-file context functions from security commits. At the training level, we propose Anchored-Contrastive Hybrid Loss (HYB), which orthogonally decouples global scale calibration from within-group defect ranking, supported by a two-pass memory-bounded gradient accumulator for long functions.

Our contribution is Cyber-JEV, a fine-tuning method in which PMC converts 17 positive-only patch archives into hash-verified groups of vulnerable, patched, and unchanged same-file functions and HYB trains these groups only through within-group ranking while keeping cross-entropy on the reference pool, yielding a five-model ensemble that outperforms Jev 1.13 on all four ranking metrics across 7,270 functions and ranks vulnerable code above its patched version more often than reference-pool fine-tuning on three patch suites, including two from TitanVul.

\section{Structured Contrastive Dataset Curation and Sanitization}
\label{sec:data}

Public Common Vulnerabilities and Exposures (CVE) records and patch repositories offer diverse examples of real-world defects, yet most extract only vulnerable functions alongside truncated spans and near-duplicate clones. PMC transforms heterogeneous raw archives into clean contrastive supervision through five sequential stages.

\subsection{Stage 1: Multi-Source Collection and Cross-Split Hygiene}

Starting from 148,462 candidate functions, Stage 1 aggregates two complementary source categories: (i) a natural reference pool $\mathcal{R}$ drawn from three benchmarks containing both natural positives and negatives---PrimeVul~\cite{ding2025primevul}, DiverseVul~\cite{chen2023diversevul}, and CVEfixes~\cite{bhandari2021cvefixes}; and (ii) 17 external security patch archives $\mathcal{T}$---including MoreFixes~\cite{akhoundali2024morefixes}, MegaVul~\cite{ni2024megavul}, ARVO~\cite{mei2024arvo}, CrossVul~\cite{nikitopoulos2021crossvul}, ReposVul~\cite{wang2024reposvul}, and curated National Vulnerability Database (NVD) commit collections.

To uphold strict evaluation hygiene, we build an undirected association graph across all training candidates and protected validation, calibration, and test splits: two functions are linked if they share a normalized open-source project family, CVE identifier, Git fix commit hash, exact code hash (under raw, whitespace-normalized, or comment-stripped lexical forms), or 128-bit token-shingle Jaccard similarity of at least 0.8, and only connected components completely disjoint from evaluation splits are admitted to training. After deduplication, reference pool $\mathcal{R}$ contains 197 vulnerable positives and 5,026 benign negatives, while the 17 external archives $\mathcal{T}$ contribute 3,357 vulnerable positives and zero negatives (top two rows of Table~\ref{tab:corpus}). Naively merging $\mathcal{R}$ and $\mathcal{T}$ causes the normalized mutual information (NMI) between source identity and label, $I(S;Y)/H(Y)$, to reach 87.3\%.

\subsection{Stages 2 to 5: Provenance-Matched Completion and Cleaning}

To eliminate source predictability and supply fine-grained repair contrasts, PMC executes four reconstruction and sanitization stages:

\emph{Stage 2: Cryptographic Git Verification and Zero-Truncation Parsing.} Raw commit datasets frequently suffer from missing file snapshots or broken spans. For each security commit, PMC verifies both the Git object \texttt{blob SHA-1} and file content \texttt{SHA-256} of pre-fix and post-fix files, then applies pinned \texttt{tree-sitter} grammars to extract complete, syntax-error-free function definitions (150 to 24,000 bytes) that fit entirely within the prompt window without token truncation. For the 725 positives that do not yield a verified non-empty group (including 660 MoreFixes entries that lack post-fix code), we enforce strict completeness gating: only the 2,829 positives that form verified non-empty contrast groups enter the contrastive ranking branch of HYB, whereas all 3,554 positives remain retained in full-data cross-entropy baselines so baselines receive maximum data volume. The 45 reference-pool positives without a group still enter the cross-entropy term of HYB, so HYB trains on 2,874 distinct positives (Table~\ref{tab:coverage}).

\begin{table}[t]
\caption{\normalfont Vulnerable Functions Used in Training}
\label{tab:coverage}
\centering
\footnotesize
\begin{tabular*}{\columnwidth}{@{\extracolsep{\fill}}lrrrr@{}}
\toprule
Source & Admitted & In a group & No group & Used\\
\midrule
Reference data & 197 & 152 & 45 & 197\\
External records & 3,357 & 2,677 & 680 & 2,677\\
Total & 3,554 & 2,829 & 725 & 2,874\\
\bottomrule
\end{tabular*}
\par\vspace{3pt}
\parbox{\columnwidth}{\footnotesize Used: distinct vulnerable functions Cyber-JEV trains on.}
\end{table}

\emph{Stage 3: Patched Version and Same-File Context Pairing.} For each verified vulnerable function $x^{+}$, PMC constructs a provenance-matched contrast group $G(x^{+})$ (2,829 groups totaling 10,099 members, median size 4) comprising two complementary types of same-source negatives:
\begin{enumerate}
\item Patched version $x^{\text{patch}}$ (present in 2,789 groups, combining 2,637 newly extracted counterparts and 152 reference-pool pairs; the remaining 40 groups pair $x^{+}$ solely with verified same-file functions): the repaired version of the same function after the commit, isolating the minimal edit that resolves the vulnerability;
\item Unchanged same-file context functions $\{c_1, \dots, c_K\}$ (deterministically sampled up to $K=2$ per positive using a hash seed keyed to sample identity, or all available if fewer than two exist, totaling 4,481 functions): other functions in the same source file that remain byte-identical across the fix commit. Sharing headers, macros, and project style, they populate external archives with authentic benign context.
\end{enumerate}

\emph{Stage 4: Near-Clone Conflict Sanitization for Weak Negatives.} Because commit-derived negatives carry weak supervision, PMC screens every candidate patched and context function against all known vulnerable functions across the entire corpus via three-form SHA-256 matching and near-clone detection (Jaccard $\ge 0.8$), discarding any candidate matching a known defect (apart from a patched function's own pre-fix counterpart). Furthermore, because external matched negatives enter HYB exclusively through within-group relative ranking rather than hard binary cross-entropy, rare unflagged minor flaws do not distort global calibration provided the actively patched defect carries higher acute risk.

\emph{Stage 5: Pre-Training Bias Audit and Held-Out Probe.} As shown in Table~\ref{tab:corpus}, pairing patched versions reduces source NMI from 87.3\% to 31.5\%, and adding $K=2$ unchanged same-file functions lowers it to 14.0\% with a near-neutral length-only AUROC of 0.574. We also hold out 1,924 never-trained unchanged same-file functions from external archives to audit cross-source false alarms.

\begin{table}[t]
\caption{\normalfont Dataset Composition and Source Predictability}
\label{tab:corpus}
\centering
\footnotesize
\setlength{\tabcolsep}{2.0pt}
\begin{tabular*}{\columnwidth}{@{\extracolsep{\fill}}lrrrrr@{}}
\toprule
Curation variant & Total & Pos. & Neg. & NMI & Len.\ AUROC \\
\midrule
Naive mix ($\mathcal{R} \cup \mathcal{T}$) & 8,580 & 3,554 & 5,026 & 87.3\% & 0.525 \\
Reference pool ($\mathcal{R}$) & 5,223 & 197 & 5,026 & 11.8\% & 0.552 \\
PMC: patched only & 11,217 & 3,554 & 7,663 & 31.5\% & 0.508 \\
PMC: patched + 2 in-file & 15,698 & 3,554 & 12,144 & 14.0\% & 0.574 \\
PMC: patched + 4 in-file & 19,403 & 3,554 & 15,849 & 9.2\% & 0.600 \\
PMC: patched + all in-file & 55,100 & 3,554 & 51,546 & 3.9\% & 0.648 \\
\midrule
Main test benchmark & 7,270 & 322 & 6,948 & 4.1\% & 0.675 \\
\bottomrule
\end{tabular*}
\par\vspace{3pt}
\parbox{\columnwidth}{\footnotesize Len.\ AUROC: ranking by function length alone.}
\end{table}

\section{Decoupled Anchored-Contrastive Training Framework}
\label{sec:method}

Even with verified matched groups, standard loss functions struggle to jointly maintain global cross-project calibration and local patch sensitivity. We analyze this conflict and present our hybrid objective and memory-bounded gradient accumulator.

\subsection{Conflict Between Global Calibration and Local Sensitivity}

Because repository formatting, macros, and naming styles expose provenance $S(X)$, Bayes' rule decomposes the optimal training log-odds (logit) into two additive terms:
\begin{equation}
\begin{aligned}
\operatorname{logit} P_{\text{train}}(Y{=}1 \mid X) &= \underbrace{\operatorname{logit} P_{\text{train}}(Y{=}1 \mid S(X))}_{\alpha(S(X)):\ \text{source style prior}} \\
&\quad + \underbrace{\log \frac{p(X \mid Y{=}1, S(X))}{p(X \mid Y{=}0, S(X))}}_{\beta(X):\ \text{within-source defect contrast}}.
\end{aligned}
\label{eq:decomp}
\end{equation}
Equation~\eqref{eq:decomp} explains why single-objective training fails at both ends. In uncompleted positive-only archives where $P_{\text{train}}(Y{=}0 \mid S(X))=0$, the source prior $\alpha(S(X))$ diverges and overwhelms $\beta(X)$, injecting severe ranking noise on unseen test repositories. Even after PMC completion, applying standard cross-entropy across all rows forces the model to fit artificial group construction ratios (roughly $1{:}3$ versus 4.4\% natural prevalence), distorting global calibration; conversely, training solely with within-group ranking learns $\beta(X)$ but discards cross-repository baseline scale, leaving the model unable to compare a risky parser function in one project against a simple logging helper in another.

\subsection{Anchored-Contrastive Hybrid Loss}

To resolve this dilemma, we combine the statistical principle of canceling group-specific repository style bias via conditional likelihood~\cite{neyman1948incidental,andersen1970cml} and anchoring baseline log-odds on a representative reference sample~\cite{prentice1979casecontrol,scott1986choice} with InfoNCE contrastive ranking~\cite{oord2018cpc,robinson2021hard}.

Let $s_{\theta}(x) = \ell_{\text{vuln}}(x) - \ell_{\text{benign}}(x)$ denote the scalar decision margin output by the classification head for function $x$. For a matched contrast group $G = \{x_0, x_1, \dots, x_m\}$ with vulnerable function $x_0 = x^{+}$ and matched negatives, we define the within-group contrastive ranking loss at temperature $\tau$ (fixed to $\tau = 1.0$ throughout):
\begin{equation}
\mathcal{L}_{\text{rank}}(G) = \log \sum_{j=0}^{m} \exp\!\big(s_{\theta}(x_j)/\tau\big) - s_{\theta}(x_0)/\tau .
\label{eq:rank}
\end{equation}
Crucially, $\mathcal{L}_{\text{rank}}(G)$ satisfies exact shift invariance to shared source biases: if $s_{\theta}(x) = a_{\theta}(z) + g_{\theta}(x)$, where $a_{\theta}(z)$ captures any bias from repository or file context $z$, then because all members of $G$ come from the exact same commit and file, the shared factor $\exp(a_{\theta}(z)/\tau)$ cancels identically between numerator and denominator in~\eqref{eq:rank}. Memorizing a project-style offset shared by the group therefore yields no reduction in this loss, pushing the shared encoder toward within-group defect distinctions $g_{\theta}(x)$.

Denoting the reference-pool rows and matched groups sampled in an optimization step by $\mathcal{R}_{\text{step}}$ and $\mathcal{G}_{\text{step}}$, HYB orthogonally combines global anchoring with local ranking:
\begin{equation}
\mathcal{L}_{\text{HYB}} = \frac{\sum_{i \in \mathcal{R}_{\text{step}}} w_i \, \mathrm{CE}_i}{\sum_{i \in \mathcal{R}_{\text{step}}} w_i} + \frac{\lambda}{|\mathcal{G}_{\text{step}}|} \sum_{G \in \mathcal{G}_{\text{step}}} \mathcal{L}_{\text{rank}}(G),
\label{eq:hyb}
\end{equation}
where the first term computes class-balanced binary cross-entropy $\mathrm{CE}_i$ exclusively on natural reference pool $\mathcal{R}$ with fixed weights $(w_1, w_0) = (N_{\text{neg}}/N_{\text{pos}}, 1)$ so positive and negative classes carry equal total weight (shifting Bayes-optimal logits only by an order-preserving constant $\log(w_1/w_0)$), while the second term distributes all 2,829 verified contrast groups from $\mathcal{R} \cup \mathcal{T}$ evenly across steps (default $\lambda = 1.0$). Because external archives $\mathcal{T}$ never enter global cross-entropy, HYB acquires fine-grained patch sensitivity without importing source-prior skew.

\subsection{Memory-Bounded Two-Pass Group Gradient Accumulation}

Complex vulnerable functions often span thousands of tokens. Evaluating~\eqref{eq:rank} requires forward scores for all members of $G$; for 4B-parameter models, roughly 7\% of long-function groups exceed single-microbatch GPU memory.

To avoid truncating code or dropping long groups, we pair a group-preserving batcher with a two-pass gradient accumulator inspired by GradCache~\cite{gao2021gradcache}. By the chain rule, the group loss gradient decomposes into a weighted sum of individual member gradients, $\nabla_{\theta} \mathcal{L}_{\text{rank}}(G) = \sum_{j} c_j \nabla_{\theta} s_{\theta}(x_j)$ with $c_j = \partial \mathcal{L}_{\text{rank}}(G)/\partial s_{\theta}(x_j)$. Pass one evaluates oversized groups across microbatches without storing activation graphs to compute the scalar softmax coefficients $c_j$; pass two re-forwards each microbatch and backpropagates $c_j s_{\theta}(x_j)$, enabling zero-truncation group training within standard single-GPU memory limits. Because pass one runs in evaluation mode and pass two in training mode with LoRA dropout, this path approximates the exact group gradient.

\begin{table*}[t]
\caption{\normalfont Ranking Results on the Main Test Benchmark}
\label{tab:main}
\centering
\footnotesize
\begin{tabular*}{\textwidth}{@{\extracolsep{\fill}}lcccccccc@{}}
\toprule
System or training variant & Models & AUROC & AP & R@1\% & R@5\% & Length AUROC & T271 pairs\\
\midrule
\multicolumn{8}{@{}l}{\textit{Reference systems}}\\
Jev 1.13, official API & -- & 0.743 & 0.101 & 2.2\% & 14.3\% & 0.706 & 50.7\%\\
Token length & -- & 0.688 & 0.125 & 7.8\% & 23.6\% & 0.558 & 16.6\%\\
Length and source tags, fitted on development and calibration data & -- & 0.734 & 0.137 & 7.8\% & 25.5\% & 0.615 & 16.6\%\\
Qwen3.5-4B without fine-tuning & 1 & 0.640 & 0.081 & 2.2\% & 14.0\% & 0.580 & 39.5\%\\
\midrule
\multicolumn{8}{@{}l}{\textit{Fine-tuned on the same backbone}}\\
Reference data and external records, classification (2 epochs) & 1 & 0.602 & 0.069 & 2.2\% & 11.2\% & 0.563 & 40.6\%\\
Reference data only, classification & 5 & 0.804 & 0.178 & 8.7\% & \underline{31.4\%} & 0.767 & 32.8\%\\
All group data, classification & 3 & 0.784 & 0.146 & 8.4\% & 25.8\% & 0.731 & \underline{55.7\%}\\
All group data, classification and ranking (2 epochs) & 1 & 0.700 & 0.101 & 3.7\% & 18.3\% & 0.635 & \textbf{61.3\%}\\
Source logit adjustment & 3 & \textbf{0.825} & \underline{0.196} & \textbf{11.2\%} & \textbf{32.0\%} & \textbf{0.784} & 28.4\%\\
Randomized group labels & 5 & 0.776 & 0.156 & 8.4\% & 26.1\% & 0.749 & 31.7\%\\
\midrule
\multicolumn{8}{@{}l}{\textit{Cyber-JEV}}\\
Five-model ensemble, main system & 5 & 0.814 & \textbf{0.197} & \underline{10.2\%} & 30.1\% & 0.768 & 47.6\%\\
Selected single model & 1 & \underline{0.820} & 0.194 & 8.7\% & \textbf{32.0\%} & \underline{0.775} & 43.9\%\\
\bottomrule
\end{tabular*}
\par\vspace{3pt}
\parbox{\textwidth}{\footnotesize Models: number of seeds whose scores are averaged; fine-tuned runs use three epochs
unless stated. R@1\%, R@5\%: recall at a test FPR of 1\% and 5\%. Length AUROC: AUROC within token-length quintiles.
T271 pairs: share of the 271 vulnerable--patched pairs in which the vulnerable version scores higher, ties counted
half. Bold: best; underlined: second best.}
\end{table*}

\begin{figure*}[t]
\centering
\includegraphics[width=\textwidth]{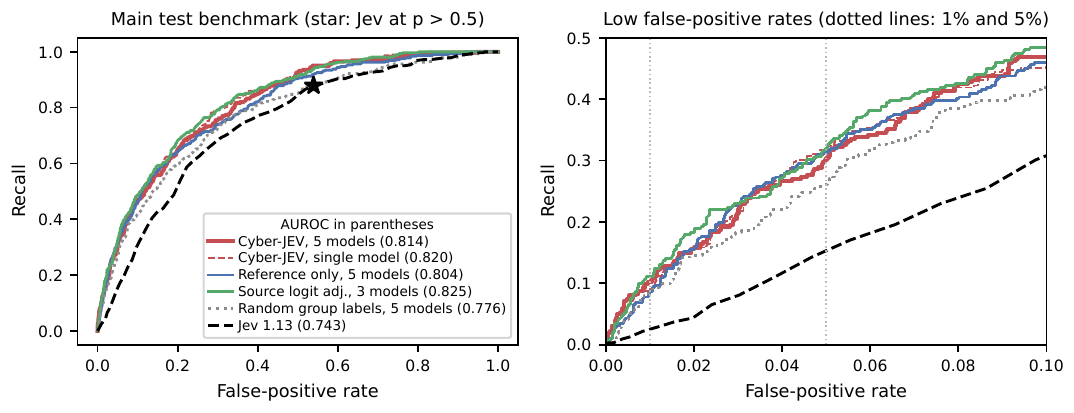}
\caption{ROC curves on the main test benchmark. Left: full range, with the default operating point of Jev marked by a
star. Right: false-positive rates up to 10\%, where a small review budget operates.}
\label{fig:roc}
\end{figure*}

\section{Experimental Setup}
\label{sec:setup}

\subsection{Model Configuration and Multi-Seed Ensemble Interface}

We fine-tune a frozen bf16 Qwen3.5-4B~\cite{qwen35report} backbone using low-rank adaptation (LoRA)~\cite{hu2022lora} (rank $r=16$, $\alpha=32$, training dropout 0.05) and a token-initialized binary classification head. All primary variants share an identical optimizer, cosine learning-rate schedule, and three-epoch budget (about 64 functions per step).

Hyperparameter locking and threshold calibration use an independent 512-function validation and calibration split with zero test-label exposure. To stabilize single-run initialization variance on compact low-prevalence validation splits, standard Cyber-JEV averages decision margins across five independent random-seed runs (individual seed AUROCs span 0.795--0.820, mean 0.805), alongside reporting the validation-selected single checkpoint (0.820 AUROC).

\subsection{Benchmarks, Operating Regimes, and Statistical Protocol}

\emph{Main Benchmark and Unseen Patch Suites.} The primary benchmark contains 7,270 functions across 57 open-source projects from PrimeVul, DiverseVul, and CVEfixes (322 vulnerable functions, 4.43\% natural prevalence), reflecting realistic power-law ecosystems spanning foundational infrastructure and 52 medium/long-tail repositories, and embedding 271 vulnerability--patch pairs (T271). To test out-of-distribution generalization on unseen repositories, we evaluate two external patch suites from TitanVul~\cite{li2026titanvul}: A72 (72 pairs) and P309 (309 pairs). To quantify surface benchmark cues, we also evaluate a token-length heuristic and a logistic regression fit on validation/calibration data using token length plus benchmark subset-tag indicators (PrimeVul, DiverseVul, or CVEfixes).

\emph{Commercial Baseline and Operational Regimes.} We query Jev 1.13~\cite{typesafe2026jev,typesafe2026api} via its official API using the identical prompt; its default threshold ($p > 0.5$) triggers 4,031 alerts. We evaluate four practical false-positive rate (FPR) and recall regimes: (1) strict low-FPR audit (recall at 1\% and 5\% FPR, R@1\% and R@5\%, a ranking measure whose threshold is placed with test labels); and, without using test labels, (2) equal alert budget (matching Jev's 4,031 alerts); (3) calibration-locked screening (threshold locked on 245 calibration negatives for $\text{FPR} \le 0.50$); and (4) high-recall noise reduction (threshold aligned to Jev's recall on validation/calibration positives). Statistical inference uses 2,000-resample 95\% project-cluster bootstrap confidence intervals (CIs) across the 57 projects and exact McNemar tests on patch pairs.

\section{Comparison with Jev and Error Analysis}
\label{sec:jev}

\subsection{Head-to-Head Triage Performance Across Operational Regimes}

Table~\ref{tab:main} and Fig.~\ref{fig:roc} summarize main benchmark performance. Zero-shot Qwen3.5-4B reaches only 0.640 AUROC, confirming that general code pretraining alone is inadequate for vulnerability triage. Cyber-JEV significantly outperforms Jev 1.13 across all four primary ranking metrics (all 95\% project-cluster CIs exclude zero): AUROC improves from 0.743 to 0.814, average precision (AP) nearly doubles from 0.101 to 0.197 (+95\% relative), and length-stratified AUROC rises from 0.706 to 0.768. Crucially, while biased ablations in the middle block of Table~\ref{tab:main} (such as source logit adjustment at 0.825 AUROC) achieve high macroscopic scores, they collapse on T271 patch pairs (28.4\%); Cyber-JEV combines strong global ranking with improved patch sensitivity.

Table~\ref{tab:confusion} translates these metrics into exact true positives (TP), missed defects (false negatives, FN), false alarms (false positives, FP), and true negatives (TN) across all 322 vulnerable and 6,948 benign test functions. At a 1\% FPR cap, Cyber-JEV catches 33 vulnerabilities versus 7 for Jev (4.6$\times$ the recall; tied scores leave Jev at 58 of the 69 allowed false alarms); under Jev's 4,031-alert budget, it cuts missed vulnerabilities from 39 to 16; at its calibration-locked threshold, it finds 18 more defects with 131 fewer false alarms than Jev's default point (Fig.~\ref{fig:roc}, left); and when aligning recall on validation/calibration positives, it cuts false alarms from 3,748 to 2,466 (34.2\%) at 81.7\% test recall versus 87.9\% for Jev, a recall difference that is not statistically significant, whereas reference-pool-only training achieves no significant FP reduction.

\begin{table}[t]
\caption{\normalfont Review Outcomes at Fixed Operating Points}
\label{tab:confusion}
\centering
\footnotesize
\setlength{\tabcolsep}{3pt}
\begin{tabular*}{\columnwidth}{@{\extracolsep{\fill}}llrrrr@{}}
\toprule
Setting & System & TP & FN & FP & TN\\
\midrule
Test FPR $\le$ 1\% & Jev 1.13 & 7 & 315 & 58 & 6,890\\
 & Cyber-JEV & 33 & 289 & 69 & 6,879\\
\midrule
Test FPR $\le$ 5\% & Jev 1.13 & 46 & 276 & 325 & 6,623\\
 & Cyber-JEV & 97 & 225 & 347 & 6,601\\
\midrule
Default, 4,031 alerts & Jev 1.13 & 283 & 39 & 3,748 & 3,200\\
Same 4,031 alerts & Cyber-JEV & 306 & 16 & 3,725 & 3,223\\
Calibrated threshold & Cyber-JEV & 301 & 21 & 3,617 & 3,331\\
\midrule
Recall target & Reference only & 264 & 58 & 2,770 & 4,178\\
 & Cyber-JEV & 263 & 59 & 2,466 & 4,482\\
\bottomrule
\end{tabular*}
\par\vspace{3pt}
\parbox{\columnwidth}{\footnotesize TP: vulnerable functions flagged; FN: missed; FP: false alarms; TN: correctly not
flagged; 322 vulnerable and 6,948 non-vulnerable functions.}
\end{table}

\subsection{Root-Cause Analysis of False Positives and False Negatives}

\begin{figure}[t]
\centering
\includegraphics[width=\columnwidth]{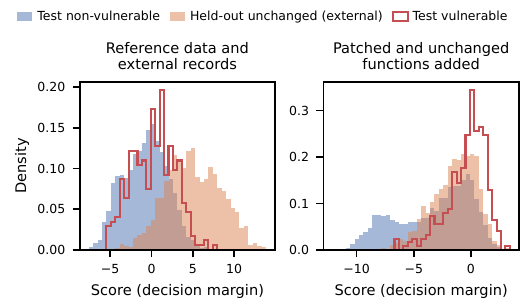}
\caption{Scores of two classification-only models, without (left) and with (right) patched and unchanged functions.
Orange: 1,924 never-trained external unchanged functions; source separation 0.916 (left) and 0.652 (right).}
\label{fig:prov}
\end{figure}

Analyzing model predictions and representation probes identifies three core mechanisms behind false positives (FP) and two likely sources of false negatives (FN):

\emph{1) FP Driver 1: Repository Style and File-Context Spillover.} In Fig.~\ref{fig:prov} (left), naive uncompleted mixing allows a linear probe to decode source identity from last-layer embeddings at 0.869 AUROC, causing the model to falsely flag 94.0\% of the 1,924 held-out benign external same-file functions. Pairing patched versions without unchanged same-file functions still leaves this cross-source FPR at 89.9\%. Full PMC and HYB cut cross-source FPR to 41.5\%, and adding patched and unchanged functions moves the held-out functions toward the benchmark negatives (Fig.~\ref{fig:prov}, right), suppressing false alarms caused by unfamiliar project conventions or co-location in risky files.

\emph{2) FP Driver 2: Function Length and Complexity Bias.} Table~\ref{tab:main} shows that logistic regression on token length and subset tags reaches 0.734 AUROC overall yet drops to 0.615 within length quintiles. Similarly, Jev 1.13 drops from 0.743 to 0.706 within length strata, consistent with a bias toward long functions. Cyber-JEV sustains 0.768 length-stratified AUROC, curbing complexity-driven false alarms.

\emph{3) FP Driver 3: Residual High-Risk Context and Score Ties on Patched Code.} Under permissive high-recall binary thresholds ($>85\%$ recall), because security patches alter only a few lines while leaving the surrounding pointer arithmetic and buffer operations unchanged, any coarse binary threshold flags both pre-fix and post-fix versions as belonging to high-risk modules. Crucially, however, Jev 1.13 outputs exact tied probabilities on 18.8\% (51/271) of T271 pairs---its apparent 50.7\% score includes 9.4 percentage points of half-credit from ties, corresponding to a strict win rate of only 41.3\% (112/271). Cyber-JEV produces continuous tie-free margins achieving a 47.6\% strict win rate (+6.3 points over Jev), ranking unpatched defects ahead of repaired code in priority queues.

\emph{4) FN Drivers and Single-Function Screening Boundaries.} The 16 missed defects at equal alert budget (5.0\% FN) and marginal FN increases under noise reduction have not been inspected one by one; likely sources are (i) inter-procedural vulnerabilities governed by caller argument validation or cross-function memory lifecycles invisible within a single function body, and (ii) compact logic flaws in short functions lacking canonical memory APIs, which receive conservative scores under mild score--length correlation (Spearman $\rho = 0.46$).

\section{Shortcut Diagnostics and Component Ablations}
\label{sec:ablation}

Table~\ref{tab:diag} jointly evaluates cross-source negative separation (AUROC distinguishing held-out external benign functions from benchmark negatives, where 0.500 is ideal; notably, an unlabeled version computed against all pooled benchmark functions matches the labeled metric within 0.015 across all models, offering a zero-label pre-deployment check) and patch-pair ranking accuracy on T271 and unseen TitanVul suites A72 and P309; length-stratified AUROC is reported in Table~\ref{tab:main}.

\begin{table}[t]
\caption{\normalfont Source Separation and Vulnerable--Patched Ranking}
\label{tab:diag}
\centering
\footnotesize
\setlength{\tabcolsep}{2.5pt}
\begin{tabular*}{\columnwidth}{@{\extracolsep{\fill}}lccccc@{}}
\toprule
System or training variant & Models & Source & T271 & A72 & P309\\
\midrule
Jev 1.13 & -- & -- & 50.7\% & -- & --\\
Qwen3.5-4B, no fine-tuning & 1 & 0.410 & 39.5\% & 22.2\% & 34.0\%\\
\multicolumn{6}{@{}l}{\textit{Classification only}}\\
Reference + external records & 1 & 0.916 & 40.6\% & 50.0\% & 44.7\%\\
Reference data only & 5 & 0.484 & 32.8\% & 33.3\% & 37.5\%\\
Source logit adjustment & 3 & 0.593 & 28.4\% & 23.6\% & 36.6\%\\
All group data & 3 & 0.669 & \textbf{55.7\%} & \underline{76.4\%} & 46.0\%\\
\multicolumn{6}{@{}l}{\textit{Cyber-JEV objective, groups varied}}\\
Randomized group labels & 5 & 0.507 & 31.7\% & 36.1\% & 38.2\%\\
Unchanged functions only & 3 & 0.419 & 35.4\% & 34.7\% & 35.6\%\\
Patched versions only & 3 & 0.523 & \textbf{55.7\%} & \textbf{80.6\%} & \textbf{63.1\%}\\
Full groups (Cyber-JEV) & 5 & 0.433 & 47.6\% & 62.5\% & \underline{52.4\%}\\
\bottomrule
\end{tabular*}
\par\vspace{3pt}
\parbox{\columnwidth}{\footnotesize Source: cross-source negative separation (0.5 means none). Pairs: vulnerable version
scored higher, ties counted half. Unchanged-only and patched-only variants have 2,330 and 2,789 groups. Pair columns
mark the best in bold and the second best underlined.}
\end{table}

\emph{Overcoming Anti-Patch Bias.} Baselines lacking patch contrasts systematically invert minimal fixes: reference-pool fine-tuning and source logit adjustment rank the vulnerable version higher on only 32.8\% and 28.4\% of T271 pairs, and 33.3\% and 23.6\% of unseen A72 pairs (persisting even on the 51 pairs where patching does not increase length). Compared with reference-pool training, full Cyber-JEV maintains 0.814 global AUROC and 0.433 source separation while boosting pair accuracy by +14.8, +29.2, and +14.9 percentage points on T271, A72, and P309 (all $p < 10^{-3}$ via exact McNemar test).

\emph{Causal Direction of Contrastive Supervision.} The random-label group baseline preserves identical group members and compute as HYB but randomly assigns which member in each group is treated as positive (corrupting 69\% of groups). Cyber-JEV beats this baseline on all 5 out of 5 random seeds (+0.055 mean single-seed AUROC gain; 0.814 vs.\ 0.776 ensemble AUROC, hierarchical 95\% CI $[+0.019, +0.095]$), while random group labels collapse T271 accuracy to 31.7\%.

\emph{Role of Matched Negatives and Weight $\lambda$.} Removing patched versions (HYB: in-file context only) drops pair accuracy to 34.7\%--35.6\%, confirming that patched versions drive fine-grained repair sensitivity; two-member $\{x^{+}, x^{\text{patch}}\}$ groups focus conditional gradients solely on the patch edit, reaching 55.7\%, 80.6\%, and 63.1\% pair accuracy. Meanwhile, unchanged same-file functions supply benign repository context and stabilize full-data representations (lifting two-epoch full-data ranking AUROC from 0.627 to 0.700). Indiscriminate negative volume, however, dilutes repair supervision: adding all 39,402 same-file functions into cross-entropy lowers T271 accuracy to 37.6\%. Varying $\lambda$ in~\eqref{eq:hyb} traces a smooth frontier: $\lambda=0.5$ lowers T271 to 35.8\%, whereas $\lambda=2.0$ raises T271 to 52.8\% and A72 to 79.2\% at a $-0.026$ global AUROC trade-off. Default $\lambda=1.0, K=2$ balances global screening and local repair sensitivity.

\section{Use in Software Review}
\label{sec:practical}

\emph{Workflow for changing code.} For a pull request, adjacent versions or a sequence of releases, Cyber-JEV fits a
five-step loop. First, extract the functions that changed. Second, score each new version. Third, score its previous
version as a reference for the change: a rise suggests the change added risk, but the absolute score of the new version
stays in the ranking, so that an already risky function is not dropped because its score barely moved. Fourth, sort by
score and cut at the review budget, for example the number of functions a team can inspect per release. Fifth, give the
selected functions deeper checks with queries, rules, a reviewer or a review assistant.

\emph{Evidence and outlook.} The pair sets contain code before and after real historical fixes, so they measure the
sensitivity to version changes that the third step relies on. They are not pull requests in progress. We have not
deployed the loop in continuous integration, run it across consecutive releases, or used it to find new
vulnerabilities; these remain to be shown.

\emph{Efficiency.} A single model reads a function once and returns its score from one forward pass. It does not
generate an explanation token by token or call repository tools. Functions are independent, so they can be scored in
batches. When the input, model weights and preprocessing are fixed, the score of an unchanged function can be reused
from a cache across versions. The main ensemble costs five forward passes per function; the selected single model costs
one and reaches an AUROC of 0.820. We did not measure throughput against other tools.

\emph{Relation to existing tools.} CodeQL builds a database from the code, runs queries over it and reports
code-scanning alerts~\cite{github2026codeql}; its incremental analysis reuses a cached database and can restrict alerts
to lines that a pull request adds or modifies~\cite{github2026codeqlincr}. Semgrep runs rule sets and offers diff-aware
scans of the files changed in a pull request~\cite{semgrep2026ci}. Copilot code review gathers repository context,
comments on a pull request and suggests fixes, and its documentation advises adding a human
review~\cite{github2026copilot}. Incremental scanning is thus not specific to our approach. A learned score adds a
different output: one number per function that a program can sort, threshold and cap at a budget, produced by an
open-weight model that can run on local hardware, so the code need not leave the organization. Rules and review
comments explain a finding, whereas the score only orders work, which fits the role of a first stage in front of these
tools. We make no claim about relative speed or accuracy, because we did not compare the tools on the same task in the
same environment.

\section{Related Work}
\label{sec:related}

\emph{Code Vulnerability Detection and Fast Decision Models.} Graph neural networks and pretrained code encoders have been widely explored for function- and line-level vulnerability detection~\cite{zhou2019devign,chakraborty2022reveal,fu2022linevul}, while recent studies audit large language models on security reasoning~\cite{ullah2024llmsec,steenhoek2023empirical}. Empirical analyses highlight that deduplication, realistic low prevalence, and minimal patch pairs expose shortcut reliance~\cite{croft2023dataquality,ding2025primevul,risse2024limits,li2026titanvul}. Building on Jev's single-step probabilistic interface~\cite{typesafe2026jev,typesafe2026api}, Cyber-JEV resolves the data and optimization challenges of adapting open language models to source-code triage.

\emph{Shortcut Mitigation and Contrastive Code Learning.} Dataset bias and shortcut learning pervade security machine learning~\cite{arp2022dos,pendlebury2019tesseract,geirhos2020shortcut}. Distributionally robust optimization~\cite{sagawa2020groupdro}, invariant risk minimization~\cite{arjovsky2019irm}, domain-adversarial networks~\cite{ganin2016dann}, and CausalVul~\cite{rahman2024causalvul} require both classes per environment, while positive-unlabeled learning~\cite{elkan2008pu,bekker2020pu} and logit adjustment~\cite{menon2021logit} adjust class priors without teaching patch sensitivity. Unlike contrastive code models using generic in-batch negatives~\cite{liu2023contrabert,cheng2022contraflow}, Cyber-JEV unites conditional likelihood~\cite{breslow1980casecontrol,prentice1979casecontrol} with commit-matched hard-negative ranking~\cite{robinson2021hard,kaushik2020cad} while anchoring global probabilities on a clean reference pool.

\section{Conclusion}
\label{sec:conclusion}

We introduced Cyber-JEV, a matched contrastive fine-tuning framework that adapts open-weight language models to fast
single-pass source-code vulnerability triage. Provenance-Matched Completion turns positive-only vulnerability records
into hash-verified groups of vulnerable, patched, and unchanged same-file functions, and the Anchored-Contrastive Hybrid
Loss trains these groups through within-group ranking while keeping cross-entropy on a reference pool with natural
labels. On a 7,270-function benchmark, the five-model Cyber-JEV ensemble outperforms Jev 1.13 on all four ranking
measures, finds 33 instead of 7 vulnerabilities at a 1\% false-positive rate, and reduces missed vulnerabilities from
39 to 16 under the same 4,031-alert budget. It also ranks vulnerable code above its patched version 14.8 to 29.2
percentage points more often than reference-pool fine-tuning on three suites of real fixes. Because each score comes
from one forward pass over one function, Cyber-JEV can serve as a first-stage filter that orders the changed functions
of pull requests and releases before deeper analysis. Future work will evaluate this workflow on live pull requests and
consecutive releases.

\bibliographystyle{IEEEtran}
\bibliography{refs_v3}

\begin{thebibliography}{10}
\providecommand{\url}[1]{#1}
\csname url@samestyle\endcsname
\providecommand{\newblock}{\relax}
\providecommand{\bibinfo}[2]{#2}
\providecommand{\BIBentrySTDinterwordspacing}{\spaceskip=0pt\relax}
\providecommand{\BIBentryALTinterwordstretchfactor}{4}
\providecommand{\BIBentryALTinterwordspacing}{\spaceskip=\fontdimen2\font plus
\BIBentryALTinterwordstretchfactor\fontdimen3\font minus
  \fontdimen4\font\relax}
\providecommand{\BIBforeignlanguage}[2]{{%
\expandafter\ifx\csname l@#1\endcsname\relax
\typeout{** WARNING: IEEEtran.bst: No hyphenation pattern has been}%
\typeout{** loaded for the language `#1'. Using the pattern for}%
\typeout{** the default language instead.}%
\else
\language=\csname l@#1\endcsname
\fi
#2}}
\providecommand{\BIBdecl}{\relax}
\BIBdecl

\bibitem{typesafe2026jev}
\BIBentryALTinterwordspacing
D.~Almeida, ``Introducing {System One Models} \& {Jev},'' TypeSafe AI, Sep.
  2026. [Online]. Available:
  \url{https://typesafe.ai/blog/introducing-system-one-models-and-jev}
\BIBentrySTDinterwordspacing

\bibitem{typesafe2026api}
\BIBentryALTinterwordspacing
{TypeSafe AI}, ``{TypeSafe API} reference,'' Official API documentation, 2026,
  accessed October 29, 2026. [Online]. Available:
  \url{https://api.typesafe.ai/redoc}
\BIBentrySTDinterwordspacing

\bibitem{steenhoek2023empirical}
B.~Steenhoek, M.~M. Rahman, R.~Jiles, and W.~Le, ``An empirical study of deep
  learning models for vulnerability detection,'' in \emph{Proc. 45th IEEE/ACM
  Int. Conf. Softw. Eng. (ICSE)}, 2023.

\bibitem{ullah2024llmsec}
S.~Ullah, M.~Han, S.~Pujar, H.~Pearce, A.~Coskun, and G.~Stringhini, ``{LLMs}
  cannot reliably identify and reason about security vulnerabilities (yet?): A
  comprehensive evaluation, framework, and benchmarks,'' in \emph{Proc. IEEE
  Symposium on Secur. Privacy (S\&P)}, 2024.

\bibitem{risse2024limits}
N.~Risse and M.~B{\"o}hme, ``Uncovering the limits of machine learning for
  automatic vulnerability detection,'' in \emph{Proc. 33rd USENIX Security
  Symposium}, 2024.

\bibitem{allamanis2019duplication}
M.~Allamanis, ``The adverse effects of code duplication in machine learning
  models of code,'' in \emph{Proc. 2019 ACM SIGPLAN Int. Symp. New Ideas, New
  Paradigms, and Reflections on Programming and Software (Onward!)}, 2019.

\bibitem{croft2023dataquality}
R.~Croft, M.~A. Babar, and M.~M. Kholoosi, ``Data quality for software
  vulnerability datasets,'' in \emph{Proc. 45th IEEE/ACM Int. Conf. Softw. Eng.
  (ICSE)}, 2023.

\bibitem{ding2025primevul}
Y.~Ding \emph{et~al.}, ``Vulnerability detection with code language models: How
  far are we?'' in \emph{Proc. 47th IEEE/ACM Int. Conf. Softw. Eng. (ICSE)},
  2025.

\bibitem{kapoor2023leakage}
S.~Kapoor and A.~Narayanan, ``Leakage and the reproducibility crisis in
  machine-learning-based science,'' \emph{Patterns}, vol.~4, no.~9, 2023.

\bibitem{akhoundali2024morefixes}
J.~Akhoundali, S.~R. Nouri, K.~Rietveld, and O.~Gadyatskaya, ``{MoreFixes}: A
  large-scale dataset of {CVE} fix commits mined through enhanced repository
  discovery,'' in \emph{Proc. 20th Int. Conf. Predictive Models and Data
  Analytics in Softw. Eng. (PROMISE)}, 2024.

\bibitem{ni2024megavul}
C.~Ni, L.~Shen, X.~Yang, Y.~Zhu, and S.~Wang, ``{MegaVul}: A {C/C++}
  vulnerability dataset with comprehensive code representations,'' in
  \emph{Proc. 21st Int. Conf. Mining Softw. Repositories (MSR)}, 2024.

\bibitem{sagawa2020groupdro}
S.~Sagawa, P.~W. Koh, T.~B. Hashimoto, and P.~Liang, ``Distributionally robust
  neural networks for group shifts: On the importance of regularization for
  worst-case generalization,'' in \emph{Int. Conf. Learning Representations
  (ICLR)}, 2020.

\bibitem{arjovsky2019irm}
M.~Arjovsky, L.~Bottou, I.~Gulrajani, and D.~Lopez-Paz, ``Invariant risk
  minimization,'' \emph{arXiv preprint arXiv:1907.02893}, 2019.

\bibitem{menon2021logit}
A.~K. Menon, S.~Jayasumana, A.~S. Rawat, H.~Jain, A.~Veit, and S.~Kumar,
  ``Long-tail learning via logit adjustment,'' in \emph{Int. Conf. Learning
  Representations (ICLR)}, 2021.

\bibitem{breslow1980casecontrol}
N.~E. Breslow and N.~E. Day, \emph{Statistical Methods in Cancer Research,
  Volume {I}: The Analysis of Case-Control Studies}, ser. IARC Scientific
  Publications.\hskip 1em plus 0.5em minus 0.4em\relax International Agency for
  Research on Cancer, 1980, no.~32.

\bibitem{neyman1948incidental}
J.~Neyman and E.~L. Scott, ``Consistent estimates based on partially consistent
  observations,'' \emph{Econometrica}, vol.~16, no.~1, pp. 1--32, 1948.

\bibitem{andersen1970cml}
E.~B. Andersen, ``Asymptotic properties of conditional maximum-likelihood
  estimators,'' \emph{Journal of the Royal Statistical Society: Series B
  (Methodological)}, vol.~32, no.~2, pp. 283--301, 1970.

\bibitem{prentice1979casecontrol}
R.~L. Prentice and R.~Pyke, ``Logistic disease incidence models and
  case-control studies,'' \emph{Biometrika}, vol.~66, no.~3, pp. 403--411,
  1979.

\bibitem{scott1986choice}
A.~J. Scott and C.~J. Wild, ``Fitting logistic models under case-control or
  choice based sampling,'' \emph{Journal of the Royal Statistical Society:
  Series B}, vol.~48, no.~2, pp. 170--182, 1986.

\bibitem{kaushik2020cad}
D.~Kaushik, E.~Hovy, and Z.~C. Lipton, ``Learning the difference that makes a
  difference with counterfactually-augmented data,'' in \emph{Int. Conf.
  Learning Representations (ICLR)}, 2020.

\bibitem{oord2018cpc}
A.~van~den Oord, Y.~Li, and O.~Vinyals, ``Representation learning with
  contrastive predictive coding,'' \emph{arXiv preprint arXiv:1807.03748},
  2018.

\bibitem{robinson2021hard}
J.~Robinson, C.-Y. Chuang, S.~Sra, and S.~Jegelka, ``Contrastive learning with
  hard negative samples,'' in \emph{Int. Conf. Learning Representations
  (ICLR)}, 2021.

\bibitem{liu2023contrabert}
S.~Liu, B.~Wu, X.~Xie, G.~Meng, and Y.~Liu, ``{ContraBERT}: Enhancing code
  pre-trained models via contrastive learning,'' in \emph{IEEE/ACM 45th Int.
  Conf. Softw. Eng. (ICSE)}, 2023, pp. 2476--2487.

\bibitem{qwen35report}
{Qwen Team}, ``{Qwen3.5}: Towards native multimodal agents,''
  \url{https://qwen.ai/blog?id=qwen3.5}, February 2026, model: Qwen/Qwen3.5-4B.

\bibitem{chen2023diversevul}
Y.~Chen, Z.~Ding, L.~Alowain, X.~Chen, and D.~Wagner, ``{DiverseVul}: A new
  vulnerable source code dataset for deep learning based vulnerability
  detection,'' in \emph{Proc. 26th Int. Symp. Research in Attacks, Intrusions
  and Defenses (RAID)}, 2023.

\bibitem{bhandari2021cvefixes}
G.~Bhandari, A.~Naseer, and L.~Moonen, ``{CVEfixes}: Automated collection of
  vulnerabilities and their fixes from open-source software,'' in \emph{Proc.
  17th Int. Conf. Predictive Models and Data Analytics in Softw. Eng.
  (PROMISE)}, 2021.

\bibitem{mei2024arvo}
X.~Mei \emph{et~al.}, ``{ARVO}: Atlas of reproducible vulnerabilities for open
  source software,'' \emph{arXiv preprint arXiv:2408.02153}, 2024.

\bibitem{nikitopoulos2021crossvul}
G.~Nikitopoulos, K.~Dritsa, P.~Louridas, and D.~Mitropoulos, ``{CrossVul}: A
  cross-language vulnerability dataset with commit data,'' in \emph{Proc. 29th
  ACM Joint European Softw. Eng. Conference and Symposium on the Foundations of
  Softw. Eng. (ESEC/FSE)}, 2021.

\bibitem{wang2024reposvul}
X.~Wang, R.~Hu, C.~Gao, X.-C. Wen, Y.~Chen, and Q.~Liao, ``{ReposVul}: A
  repository-level high-quality vulnerability dataset,'' in \emph{Proc. 46th
  IEEE/ACM Int. Conf. Softw. Eng.: Companion Proceedings}, 2024.

\bibitem{gao2021gradcache}
L.~Gao, Y.~Zhang, J.~Han, and J.~Callan, ``Scaling deep contrastive learning
  batch size under memory limited setup,'' in \emph{Proc. 6th Workshop on
  Representation Learn. NLP (RepL4NLP)}, 2021.

\bibitem{hu2022lora}
E.~J. Hu \emph{et~al.}, ``{LoRA}: Low-rank adaptation of large language
  models,'' in \emph{Int. Conf. Learning Representations (ICLR)}, 2022.

\bibitem{li2026titanvul}
Y.~Li \emph{et~al.}, ``Out of distribution, out of luck: How well can {LLMs}
  trained on vulnerability datasets detect top 25 {CWE} weaknesses?'' in
  \emph{IEEE/ACM 48th Int. Conf. Softw. Eng. (ICSE)}, 2026, arXiv:2507.21817.

\bibitem{github2026codeql}
\BIBentryALTinterwordspacing
{GitHub}, ``Code scanning with {CodeQL},'' GitHub Docs, 2026, accessed October
  30, 2026. [Online]. Available:
  \url{https://docs.github.com/en/code-security/concepts/code-scanning/codeql/codeql-code-scanning}
\BIBentrySTDinterwordspacing

\bibitem{github2026codeqlincr}
\BIBentryALTinterwordspacing
------, ``Using incremental analysis with the {CodeQL} {CLI},'' GitHub Docs,
  2026, accessed October 30, 2026. [Online]. Available:
  \url{https://docs.github.com/en/code-security/how-tos/find-and-fix-code-vulnerabilities/scan-from-the-command-line/incremental-analysis}
\BIBentrySTDinterwordspacing

\bibitem{semgrep2026ci}
\BIBentryALTinterwordspacing
{Semgrep}, ``Sample continuous integration ({CI}) configurations,'' Semgrep
  Documentation, 2026, accessed October 30, 2026. [Online]. Available:
  \url{https://docs.semgrep.dev/semgrep-ci/sample-ci-configs}
\BIBentrySTDinterwordspacing

\bibitem{github2026copilot}
\BIBentryALTinterwordspacing
{GitHub}, ``{GitHub Copilot} code review,'' GitHub Docs, 2026, accessed October
  30, 2026. [Online]. Available:
  \url{https://docs.github.com/en/copilot/concepts/agents/code-review}
\BIBentrySTDinterwordspacing

\bibitem{zhou2019devign}
Y.~Zhou, S.~Liu, J.~Siow, X.~Du, and Y.~Liu, ``Devign: Effective vulnerability
  identification by learning comprehensive program semantics via graph neural
  networks,'' in \emph{Adv. Neural Inf. Process. Syst. (NeurIPS)}, 2019.

\bibitem{chakraborty2022reveal}
S.~Chakraborty, R.~Krishna, Y.~Ding, and B.~Ray, ``Deep learning based
  vulnerability detection: Are we there yet?'' \emph{IEEE Transactions on
  Software Engineering}, vol.~48, no.~9, 2022.

\bibitem{fu2022linevul}
M.~Fu and C.~Tantithamthavorn, ``{LineVul}: A transformer-based line-level
  vulnerability prediction,'' in \emph{Proc. 19th Int. Conf. Mining Softw.
  Repositories (MSR)}, 2022.

\bibitem{arp2022dos}
D.~Arp \emph{et~al.}, ``Dos and don'ts of machine learning in computer
  security,'' in \emph{31st USENIX Security Symposium}, 2022, pp. 3971--3988.

\bibitem{pendlebury2019tesseract}
F.~Pendlebury, F.~Pierazzi, R.~Jordaney, J.~Kinder, and L.~Cavallaro,
  ``{TESSERACT}: Eliminating experimental bias in malware classification across
  space and time,'' in \emph{28th USENIX Security Symposium}, 2019, pp.
  729--746.

\bibitem{geirhos2020shortcut}
R.~Geirhos \emph{et~al.}, ``Shortcut learning in deep neural networks,''
  \emph{Nature Machine Intelligence}, vol.~2, pp. 665--673, 2020.

\bibitem{ganin2016dann}
Y.~Ganin \emph{et~al.}, ``Domain-adversarial training of neural networks,''
  \emph{Journal of Machine Learning Research}, vol.~17, no.~59, pp. 1--35,
  2016.

\bibitem{rahman2024causalvul}
M.~M. Rahman, I.~Ceka, C.~Mao, S.~Chakraborty, B.~Ray, and W.~Le, ``Towards
  causal deep learning for vulnerability detection,'' in \emph{Proc. 46th
  IEEE/ACM Int. Conf. Softw. Eng. (ICSE)}, 2024.

\bibitem{elkan2008pu}
C.~Elkan and K.~Noto, ``Learning classifiers from only positive and unlabeled
  data,'' in \emph{ACM SIGKDD Int. Conf. Knowl. Discovery Data Mining}, 2008,
  pp. 213--220.

\bibitem{bekker2020pu}
J.~Bekker and J.~Davis, ``Learning from positive and unlabeled data: A
  survey,'' \emph{Machine Learning}, vol. 109, pp. 719--760, 2020.

\bibitem{cheng2022contraflow}
X.~Cheng, G.~Zhang, H.~Wang, and Y.~Sui, ``Path-sensitive code embedding via
  contrastive learning for software vulnerability detection,'' in \emph{ACM
  SIGSOFT Int. Symp. Software Testing and Analysis (ISSTA)}, 2022, pp.
  519--531.

\end{thebibliography}

\end{document}